\documentclass[conference,twocolumn]{IEEEtran}
\IEEEoverridecommandlockouts

\usepackage{algorithmic}
\usepackage{graphicx}
\usepackage{textcomp}
\usepackage{xcolor}
\usepackage{url}
\usepackage{tabularx}
\usepackage{adjustbox}
\usepackage{nameref}
\usepackage{longtable}
\usepackage{multirow}
\usepackage{threeparttable}
\usepackage{supertabular,booktabs}
\usepackage{colortbl}
\usepackage{cite}
\usepackage{amsmath}
\usepackage{subcaption}
\usepackage{footnote}
\usepackage{siunitx} 
\usepackage[colorlinks=true, linkcolor=blue, citecolor=blue, urlcolor=black]{hyperref}
\def\BibTeX{{\rm B\kern-.05em{\sc i\kern-.025em b}\kern-.08em
    T\kern-.1667em\lower.7ex\hbox{E}\kern-.125emX}}
\begin{document}

\title{Energy-Efficient Hardware Acceleration \\of Whisper ASR on a CGLA}


\author{
    \IEEEauthorblockN{Takuto ANDO, 
    Yu ETO Ayumu TAKEUCHI and
    Yasuhiko NAKASHIMA}
    \IEEEauthorblockA{Nara Institute of Science and Technology, 8916-5 Takayama-cho, Ikoma, Nara 630-0192, Japan.}
}


\maketitle

\thispagestyle{empty}

\begin{abstract}

    The rise of generative AI for tasks like Automatic Speech Recognition (ASR) has created a critical energy consumption challenge. While ASICs offer high efficiency, they lack the programmability to adapt to evolving algorithms. 
    To address this trade-off, we implement and evaluate Whisper's core computational kernel on the IMAX, a general-purpose Coarse-Grained Linear Arrays (CGLAs) accelerator. 
    To our knowledge, this is the first work to execute a Whisper kernel on a CGRA and compare its performance against CPUs and GPUs. 
    Using hardware/software co-design, we evaluate our system via an FPGA prototype and project performance for a 28\,nm ASIC. 
    Our results demonstrate superior energy efficiency. 
    The projected ASIC is 1.90\,$\times$ more energy-efficient than the NVIDIA Jetson AGX Orin and 9.83\,$\times$ more than an NVIDIA RTX 4090 for the Q8\_0 model. 
    This work positions CGLA as a promising platform for sustainable ASR on power-constrained edge devices.
    \end{abstract}
    

\section{Introduction}
In recent years, Artificial Intelligence~(AI) technology has undergone remarkable advancements, particularly in the field of generative AI, represented by Large Language Models~(LLMs)\nobreak\cite{expl_llm,healai,llm_mas} and Automatic Speech Recognition~(ASR) systems\nobreak\cite{ASR_1}. 
State-of-the-art~(SOTA) ASR models like Whisper\nobreak\cite{radford2022robustspeechrecognitionlargescale} have achieved human-parity recognition accuracy, leading to their rapid adoption in services such as smart assistants\nobreak\cite{ASR_smart_voice_assistant}, real-time transcription\nobreak\cite{ASR_T2T} and medical applications\nobreak\cite{ASR_emergency_medicine}.

However, this technological innovation creates a critical energy consumption challenge.
This explosive growth is driven by a heavy reliance on computational infrastructure, particularly General-Purpose Graphics Processing Units~(GPGPUs). 
The International Energy Agency~(IEA) projects that data center electricity consumption could double by 2030, reaching approximately 945 TWh\nobreak\cite{iea_energy_ai}. 
This figure is fueled primarily by the expanding demand for AI and slightly exceeds Japan's total annual electricity consumption. 
As AI models grow in complexity, relying on a single, power-intensive, general-purpose architecture is both inefficient and unsustainable.

In response to this energy crisis, specialized hardware such as Application-Specific Integrated Circuits~(ASICs) and Field-Programmable Gate Arrays~(FPGAs) have been proposed as solutions. 
Among these, the Coarse-Grained Reconfigurable Arrays~(CGRAs) are recognized as a promising architectural paradigm. 
By directly mapping dataflow graphs onto an array of processing elements~(PEs), CGRAs mitigate memory-bound operations, achieving both high throughput and energy efficiency. 
Building upon this principle, the Coarse-Grained Linear Arrays~(CGLAs) architecture, IMAX\nobreak\cite{imax_access}, is a general-purpose accelerator designed to overcome the trade-off between the power efficiency of ASICs and the programmatic flexibility of GPGPUs. 
IMAX features a unique structure that linearly arranges PEs and Local Memory Modules~(LMMs), enabling it to absorb irregular memory access patterns and deliver high throughput and power efficiency concurrently.

\begin{figure}[t]
  \centering
  \includegraphics[width=1\columnwidth]{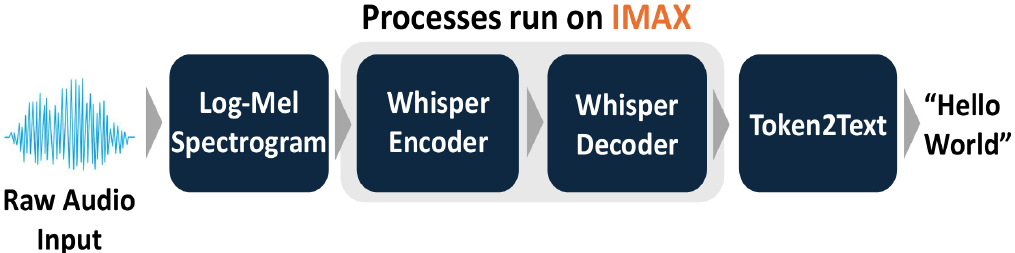}
  \caption{\small The Whisper ASR processing flow. The highlighted kernels are the computational stages accelerated by IMAX.}
  \label{fig:whisper_process}
  \vspace{-1em}

\end{figure}

\begin{figure*}[t]
  \centering
  \includegraphics[width=1.9\columnwidth]{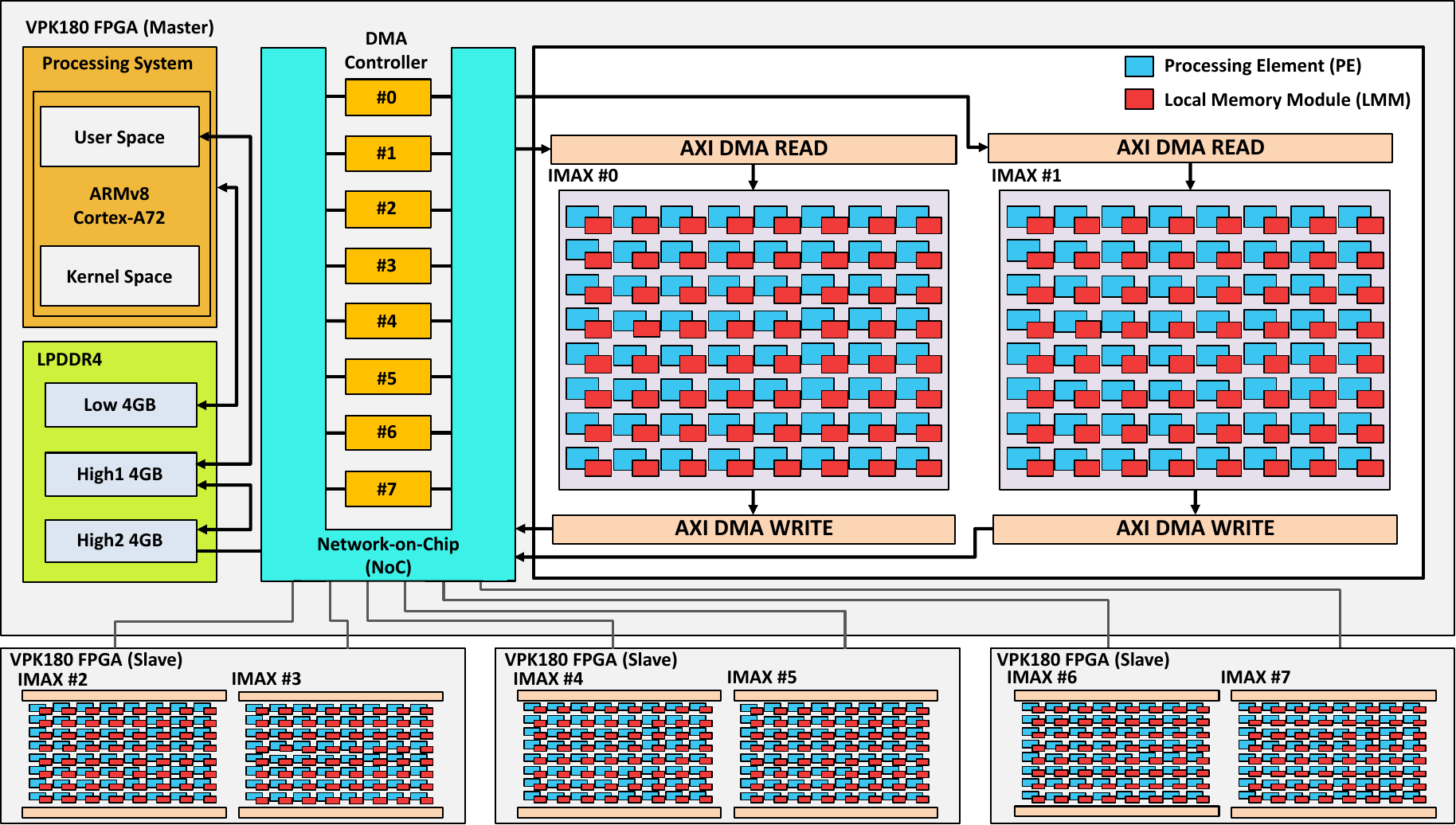}
  \caption{\small High-level overview of the IMAX3 architecture, implemented on a multi-FPGA platform with four AMD Versal VPK180 devices}
  \label{fig:imax3_conf}
\end{figure*}

Therefore, we implement and evaluate the primary computational kernels of ASR on IMAX, a CGLA accelerator. 
Our goal is to demonstrate the potential of IMAX for enabling high-performance, low-power ASR, particularly on power-constrained edge devices.
Fig.~\ref{fig:whisper_process} shows the ASR processing flow using the Whisper model. 
The encoder and decoder, which form the core of the Whisper model, consume most of the computational power. 
In this work, we focus on accelerating the dot-product kernel, which is the computational core of this model, by offloading it to IMAX.
We conduct an in-depth exploration of hardware-software co-design tailored to the unique challenges of Whisper, systematically performing kernel-level optimizations and a design space exploration of critical architectural parameters, such as LMM size.

The main contributions of this paper are as follows:
\begin{itemize}
\item We present the first implementation and evaluation of a Whisper ASR kernel on a CGRA architecture, establishing key hardware/software co-design principles for efficiently handling its dynamic and variable-length workloads.
\item Based on estimations from our FPGA prototype, we demonstrate that our optimized 28\,nm ASIC configuration has the potential to achieve superior energy efficiency, outperforming a leading-edge GPU and a high-end GPU by 1.90$\times$ and 9.83$\times$, respectively, for the Q8\_0 quantized model.
\item We analyze the trade-off between LMM size and overall energy efficiency, demonstrating that a 32\,KB LMM configuration strikes an optimal balance between maximizing kernel coverage and minimizing static power overhead.
\end{itemize}


The remainder of this paper is organized as follows. 
Section~\ref{rwork} provides an overview of related work in AI accelerators and describes the previous work on the IMAX3 architecture. 
Section~\ref{proposed} details our proposed implementation methodology. 
Section~\ref{ex_and_re} presents the experimental results and analysis. 
Section~\ref{discussion} offers further discussion, including an ablation study, and Section~\ref{conclusions} concludes the paper.

\section{Related Work}
\label{rwork}

\subsection{Hardware Accelerators for AI}
Previous work on hardware acceleration for deep learning can be broadly categorized into two approaches.
The first is a specialized approach using ASICs or FPGAs.
ASIC-based accelerators achieve extremely high performance and power efficiency by being tailored to specific neural network models or operations\nobreak\cite{llama2fpga,multi_task}.
For instance, Park et al. developed a hybrid system combining a CNN on an FPGA with a language model on a smartphone\nobreak\cite{park2022lowlatency}. 
Similarly, Hu et al. created a specialized FPGA accelerator for a GCNN model\nobreak\cite{Hu_2022}, while Yamini et al. accelerated an end-to-end Transformer-based ASR system using a systolic array\nobreak\cite{yamini2023hardware}. 
Although these studies demonstrate high performance by optimizing for specific models, their inherent inflexibility poses a significant limitation. This specialization makes it difficult to adapt to rapidly evolving algorithms and accelerates hardware obsolescence.

In contrast, the IMAX accelerator used in our work is a general-purpose architecture not tied to any specific AI task\nobreak\cite{imax_access}. 
This design enables rapid adaptation to algorithmic changes and new models. 
The novelty of this work lies in implementing Whisper, a SOTA ASR model with no prior hardware implementation examples, on this flexible, general-purpose architecture.
\subsection{CGLA Architecture and IMAX}
CGRAs are a promising architecture, balancing the efficiency of ASICs with the flexibility of GPUs. 
However, conventional CGRAs face challenges in scalability and compile time\nobreak\cite{cgra1}. 
To address these limitations, our work utilizes IMAX, specifically its latest generation, IMAX3, an architecture based on a CGLA that evolved from the CGRA concept\nobreak\cite{imax_access}.

\begin{figure}[t]
    \centering
    \includegraphics[width=0.9\columnwidth]{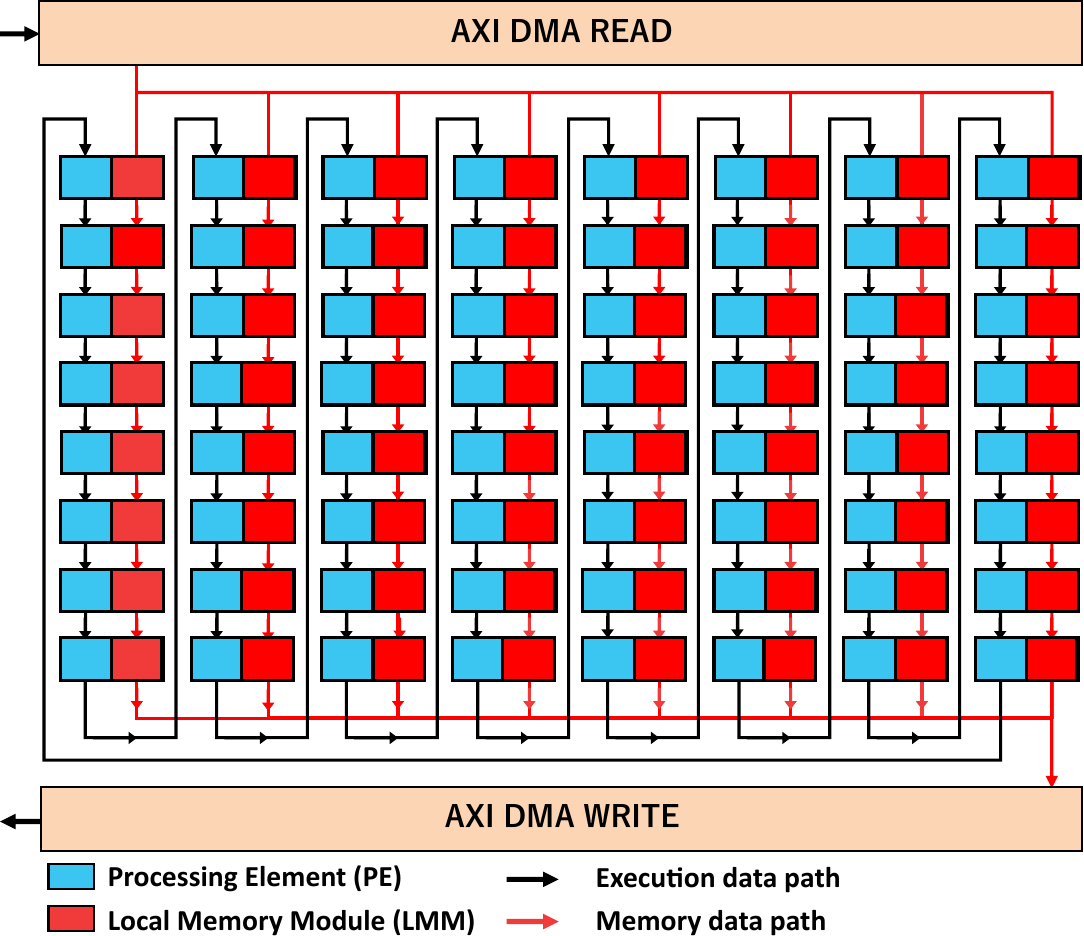}
    \caption{\small Internal structure of an IMAX lane, featuring interleaved PEs and LMMs to improve dataflow.}
    \label{fig:pe_detailed}
\end{figure}

As shown in Fig.~\ref{fig:imax3_conf}, IMAX3 is implemented in an 8-lane configuration on an AMD Versal VPK180. 
At the system level, IMAX3 is built as a System-on-Chip~(SoC). 
A Processing System~(PS) containing an ARM Cortex-A72 CPU is connected to the Programmable Logic~(PL) via a Network-on-Chip~(NoC). 
The PL hosts the CGLA cores. 
Fig.~\ref{fig:pe_detailed} illustrates a single lane of an IMAX core. 
The fundamental design of the IMAX architecture features a strategic interleaving of PEs and LMMs within a linear array structure.
This configuration effectively hides latency from irregular memory accesses, maximizes the utilization of LMM within the PEs, and alleviates memory access bottlenecks\nobreak\cite{imax_access}. 
Each PE consists of a pipelined Arithmetic Logic Unit~(ALU) and LMM, designed to maintain high processing throughput.

Our previous work has established IMAX3 as a general-purpose computation platform. 
We have demonstrated the effectiveness of IMAX3 across a wide range of workloads. 
These range from conventional compute-intensive kernels like SpGEMM and FFT\nobreak\cite{imax_access} to modern AI workloads including CNNs\nobreak\cite{unetimax,imaxcnn3}, LLMs\nobreak\cite{uetanicgra,eto2025implementation} and approximate k-NN Search\nobreak\cite{kim-knn} performed by Retrieval-Augmented Generation~(RAG). 
Building on these successes, this paper extends the applicability of IMAX3 to ASR. Specifically, we focus on the dot-product operation, which is central to this task. 
We implement and evaluate its execution on the IMAX3 architecture.

\section{Implementation of Whisper.cpp on IMAX}
\label{proposed}

This section details the hardware/software co-design we developed to enable energy-efficient execution of principal Whisper kernels on the IMAX CGLA architecture.

\subsection{Implementation Substrate and Design Objectives}
We select whisper.cpp\nobreak\cite{ggerganov_whisper.cpp} as our implementation substrate because its minimal dependencies and efficient memory management facilitate direct, low-level hardware optimization. 
This choice enables an accurate evaluation of the IMAX accelerator, avoiding the overhead of high-level frameworks. Our evaluation targets the FP16 and Q8\_0 quantized versions of the Whisper-tiny.en model.

Fully utilizing a CGRA architecture like IMAX requires a hardware/software co-design approach.
To efficiently execute a complex AI workload like Whisper, our co-design strategy has two primary objectives. 
First, we maximize computational throughput by fully leveraging IMAX's parallel processing capabilities to enhance the performance per PE. 
Second, we maximize data transfer efficiency by improving effective memory bandwidth and utilizing the LMM efficiently, which we achieve through the elimination of unnecessary data padding and compact data packing.

Our analysis of the Whisper-tiny.en model's computational profile reveals that the vast majority of the execution time is spent on dot-product operations within the multi-head attention and feed-forward network layers. 
Therefore, accelerating this specific kernel is paramount to improving the overall performance of the ASR task. 
We offload these dot-product operations, which account for the majority of its computational workload.

\subsection{Kernel-Level Co-design}

To maximize computational throughput, we introduce a new FP16 dot-product kernel while reusing the Q8\_0 kernel from the previous work\nobreak\cite{eto2025implementation}.   

Our FP16 kernel exploits IMAX's architectural features through several key optimizations.
First, we leverage the high programmability of IMAX to perform FP16-to-FP32 type conversions as an inline process, using the PE's bit manipulation capabilities to avoid dedicated hardware.
Second, we exploit two of IMAX's parallelization features to boost throughput:
\begin{itemize}
    \item SIMD Operations: We apply SIMD operations to the Fused Multiply-Add~(FMA) units, which are a primary computational bottleneck. This allows two 32-bit operations to execute concurrently on a single 64-bit datapath, enhancing throughput.
    \item Column-wise Multithreading: We employ column-wise multithreading to time-multiplex four logical FMA operations onto a single physical FPU. This technique effectively hides FPU latency and maximizes hardware resource utilization.
\end{itemize}
Third, we address the challenge of efficiently processing the variable-length vectors inherent in the Whisper workload. 
While IMAX achieves high efficiency through fixed-length burst operations, handling the residual elements of these vectors would require complex control logic such as conditional stores.
To overcome this, we introduce a mixed-execution strategy. 
This approach partitions each vector into two segments. 
The main segment, a multiple of a predefined burst length, is processed efficiently on IMAX.
The small residual segment is processed concurrently on a host CPU core, while the main segment is offloaded to IMAX. 
This hybrid strategy maximizes IMAX's efficiency on fixed-length bursts while ensuring correctness.
The effectiveness hinges on the critical trade-off in selecting an optimal burst length, a larger burst reduces overhead but also decreases the offload rate.
Based on our analysis of Whisper's vector length distribution, we determined that a burst length of 16 elements was found to be optimal in our experiments.
This choice allows us to maintain a high offload rate while still benefiting from IMAX's efficient burst operations. 
Our workload analysis confirms that the residual processing on the CPU accounts for only about \SI{5}{\percent} of the total computation.
Therefore, this mixed-execution strategy strikes an effective balance between performance and hardware simplicity.

\subsection{Optimizing Data Handling and LMM Configuration}
Maximizing data transfer efficiency and effectively utilizing the LMM are other critical components of our co-design methodology. 
The FP16 data tensors in whisper.cpp can contain significant padding to meet 32-byte alignment requirements. 
Transferring this padding wastes memory bandwidth and consumes valuable LMM capacity unnecessarily. 
To address this, we introduce a technique where the host CPU strips all padding and packs the data densely into a DMA buffer before initiating the offload.
Table~\ref{tab:padding_optimization_concise} shows the cumulative distribution of kernel memory footprints for the baseline (with padding) and our optimized implementation (without padding). 
As the table indicates, the baseline FP16 model can accommodate only \SI{1.39}{\percent} of kernels even with a 32\,KB LMM. 
In contrast, by applying our padding removal optimization, a 32\,KB LMM can cover \SI{93.80}{\percent} of all kernels, dramatically reducing on-chip memory requirements.

We determined our final LMM size based on this analysis. Table~\ref{tab:lmm_size_power_simple} presents power estimates for different LMM sizes, derived from logic synthesis using Synopsys Design Compiler with a TSMC \SI{28}{\nm} process library. 
The table shows that for an FP16 kernel, increasing the LMM size from 16\,KB to 32\,KB results in a negligible power increase from \SI{0.665}{\watt} to \SI{0.675}{\watt}. 
Meanwhile, as shown in Table~\ref{tab:padding_optimization_concise}, expanding the LMM to 32\,KB substantially increases kernel coverage from \SI{66.35}{\percent} to \SI{93.80}{\percent}. 
This improvement promises significant performance gains and dynamic power reduction by minimizing costly DRAM accesses.
Based on this analysis of the performance-power trade-off, we selected 32\,KB as the default LMM size for our evaluation. 
We will quantitatively validate the optimality of this configuration through a detailed ablation study in Section~\ref{discussion}.

\section{Experiments and Results}
\label{ex_and_re}
In this section, we evaluate the performance of the main whisper.cpp computation kernels on the IMAX3 accelerator. 
We evaluate our FPGA prototype and use logic synthesis results to project the performance of a future ASIC implementation.

\subsection{Experimental Setup}
We use the whisper.cpp framework to evaluate the transcription task on its standard jfk.wav audio file, which is approximately 10 seconds long.
Our evaluation targets the FP16 and Q8\_0 versions of the Whisper-tiny.en model.

We prototyped the IMAX accelerator on an AMD Versal Premium VPK180 evaluation kit using Vivado 2024.1. 
The SoC consists of a PS with a dual-core ARM Cortex-A72 CPU and a PL hosting an 8-lane, 64-PE IMAX computation unit. 
We operated the FPGA prototype at 140 MHz and evaluated it in one- and two-lane configurations. 
In addition to the prototype, we include performance projections for an ASIC implementation assuming an 840 MHz operating frequency on a 28\,nm process.

For comparison, we selected the embedded ARM Cortex-A72 CPU, the NVIDIA Jetson AGX Orin edge GPU in its most efficient mode, and the NVIDIA GeForce RTX 4090 high-end desktop GPU. 
Table~\ref{tab:processor_comparison_annotated} summarizes the key specifications for each platform.
We based our ASIC performance projection on physical design analysis. 
Using a TSMC \SI{28}{\nm} process library, which offers a good balance of performance and cost for edge devices, static timing analysis with Synopsys Design Compiler confirmed a critical path delay that allows for a maximum operating frequency of \SI{840}{\mega\hertz}. 
This projected 6$\times$ frequency improvement over the \SI{140}{\mega\hertz} FPGA implementation is a realistic estimate, reflecting the inherent advantages of ASIC over FPGA, such as optimized standard cell layouts, reduced routing delays, and dedicated clock distribution networks. 
We estimated the power consumption for the IMAX (ASIC) from logic synthesis results using the same Synopsys Design Compiler\cite{SynopsysNDDCUltra} and TSMC \SI{28}{\nm} library. 
For the 32\,KB LMM configuration used in our evaluation, the estimated power is 0.647\,W for FP16 kernels and 1.32\,W for Q8\_0 kernels per one-lane configuration (in case of 2 lanes, the power is 1.294\,W and 2.64\,W, respectively). 
For the comparison platforms, we used an estimated value for the host CPU and the nominal Thermal Design Power~(TDP) for the Jetson AGX Orin and GeForce RTX 4090. 
In the GPU environment, power consumption during CPU-bound phases is calculated as the sum of the host CPU (Xeon W-2455X) power\nobreak\cite{Intel_Xeon_w5-2455X_Specs} and the GPU's idle power.
Our power comparison is limited by the use of nominal TDP for the GPUs. 
Since TDP represents peak power instead of average workload power, the results should be interpreted as an indicator of architectural potential rather than a definitive measure of superiority.
A detailed analysis using measured average power would be necessary for a precise comparison, which remains as future work.
End-to-end latency was measured on the host ARM CPU using the gettimeofday function to capture wall-clock time with microsecond precision.

\begin{table}[t]
    \centering
    \caption{Cumulative percentage of each model up to the specified LMM size limit}
    \label{tab:padding_optimization_concise}
    \setlength{\tabcolsep}{7pt} 
    \begin{tabular}{l rr rr}
      \toprule
      \multirow{2}{*}{\textbf{LMM Limit}} & \multicolumn{2}{c}{\textbf{F16 Model}} & \multicolumn{2}{c}{\textbf{Q8\_0 Model}} \\ \cmidrule(lr){2-3} \cmidrule(lr){4-5}
      & \textbf{Baseline} & \textbf{Optimized} & \textbf{Baseline} & \textbf{Optimized} \\
      \midrule
      \hspace{8pt}8KB   & 0.00\%   & 64.96\% & 0.00\%   & 64.96\% \\
      \hspace{4pt}16KB  & 1.39\%   & 66.35\% & 1.39\%   & 66.35\% \\
      \hspace{4pt}\textbf{32KB}  & 1.39\%   & \textbf{93.80\%} & 28.83\%  & \textbf{93.80\%} \\
      \hspace{4pt}64KB  & 93.81\%  & 93.80\% & 93.81\%  & 93.81\% \\
      128KB & 94.49\%  & 100.00\%& 97.24\%  & 100.00\% \\
      256KB & 100.00\% & 100.00\%& 100.00\% & 100.00\% \\
      \bottomrule
    \end{tabular}
\end{table}

\begin{table}[t]
    \centering
    \caption{Power consumption by LMM size for FP16 and Q8\_0 kernels (number of units shown)}
    \label{tab:lmm_size_power_simple}
    \setlength{\tabcolsep}{5pt}
    
    \begin{tabular}{lrrrrrr}
      \toprule
      \multirow{2}{*}{\textbf{Kernel}} & \multirow{2}{*}{\textbf{Unit}} & \multicolumn{5}{c}{\textbf{LMM size}} \\
      \cmidrule(lr){3-7}
      & & \textbf{16\,KB} & \textbf{32\,KB} & \textbf{64\,KB} & \textbf{128\,KB} & \textbf{256\,KB} \\
      \midrule
      FP16   & 22 & 0.637\,W &\textbf{0.647\,W} & 2.16\,W & 5.18\,W & 11.2\,W \\
      Q8\_0  & 46 & 1.3\,W   &\textbf{1.32\,W}  & 4.41\,W & 10.6\,W & 22.9\,W \\
      \bottomrule
    \end{tabular}
    \vspace{-1.5em}
  \end{table}

\begin{table*}[t]
    \centering
    \begin{threeparttable}
    
    \caption{Physical specifications and performance comparisons of evaluated hardware platforms}
    \label{tab:processor_comparison_annotated}
    
    \begin{tabular}{@{} llrrrrrrr l @{}} 
      \toprule
      \textbf{Device} & \textbf{CPU} & \textbf{Cores} & \textbf{Chip area} & \textbf{Process} & \textbf{Operating frequency} & \textbf{Memory} & \textbf{Power}\tnote{c} \\
      & & & \textbf{(mm$^2$)} & \textbf{node} & & & \textbf{(W)} \\
      \midrule
   
      \textbf{ARM Cortex-A72 (on Versal)} & - & 2 & - & 7\,nm & 1400\,MHz & 8\,GB DDR4 & 0.6485 \\
   
      \textbf{IMAX3 (Xilinx VPK180)} & ARM Cortex-A72 & 64\tnote{a} & - & 7\,nm & 145\,MHz & 8\, + 4\,GB DDR4\tnote{b} & 180 \\
   
      \textbf{IMAX3 (28nm)} & - & 64\tnote{a} & 14.6 & \SI{28}{\nano\meter} & 840\,MHz& - &0.647 or 1.32 \\
   
      \textbf{Jetson AGX Orin 32GB} & Arm® Cortex-A78AE & 1792 & 200 & 8\,nm & 930\,MHz & 32\,GB DDR5 & 15\tnote{d} \\
   
      \textbf{NVIDIA RTX 4090} & Xeon W5-2455X & 16384 & 608 & 5\,nm & 2520\,MHz & 24\,GB DDR6 & 450 \\
      \bottomrule
    \end{tabular}
    
  
   \begin{tablenotes}[para,flushleft]
     \item[a] The number of cores for IMAX3 refers to the number of PEs per lane.
     \item[b] 8\,GB DDR4 for OS buffer and 4\,GB DDR4 for DMA buffer.
     \item[c] IMAX3~(28\,nm) is an estimated value, references for other devices are from Cortex-A72\nobreak\cite{Versal}, Jetson AGX Orin 32GB\nobreak\cite{nvidia_jetson_agx_orin}, NVIDIA RTX 4090\nobreak\cite{nvidia_ada}.
     \item[d] This device has multiple power consumption modes~(15--45\,W), and we executed in the lowest power consumption mode.
   \end{tablenotes}
    
     \end{threeparttable}
    \end{table*}

\begin{figure}[t]
    \centering
    
    \begin{subfigure}{1.0\columnwidth}
      \centering
      \includegraphics[width=\textwidth]{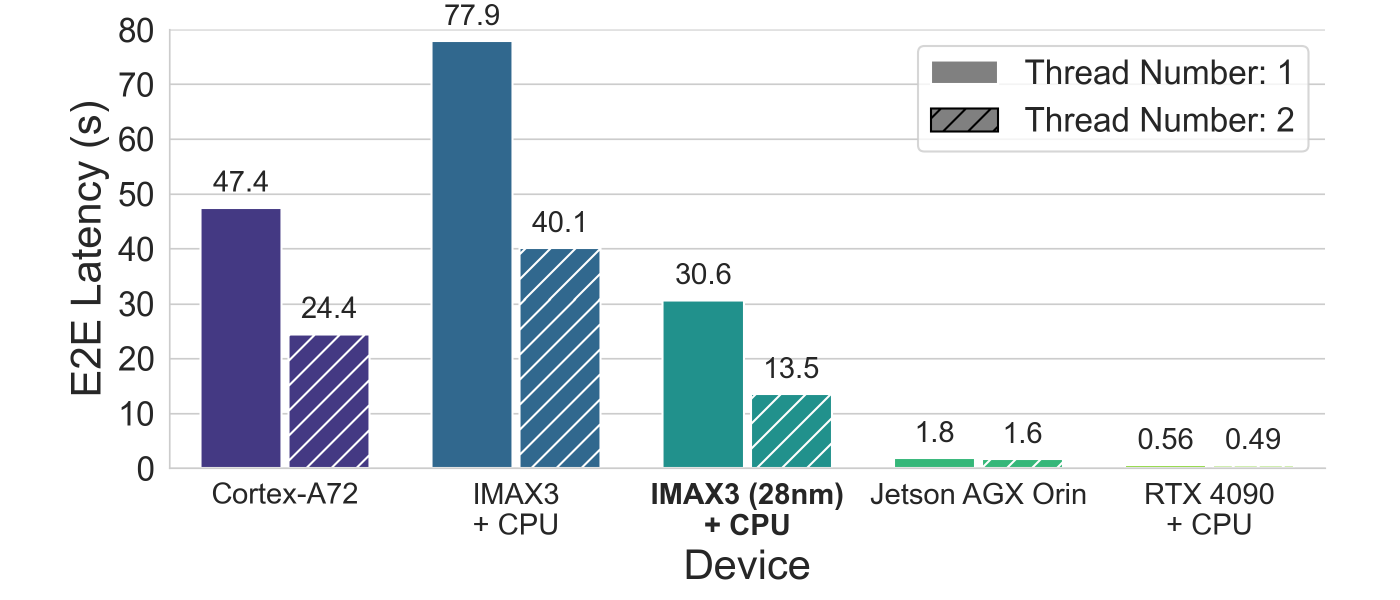}
      \subcaption{\small Whisper-tiny.en FP16 model inference}
      \label{fig:e2e_FP16}
    \end{subfigure}
    
    \vspace{6pt} 
    
    \begin{subfigure}{1.0\columnwidth}
      \centering
      \includegraphics[width=\textwidth]{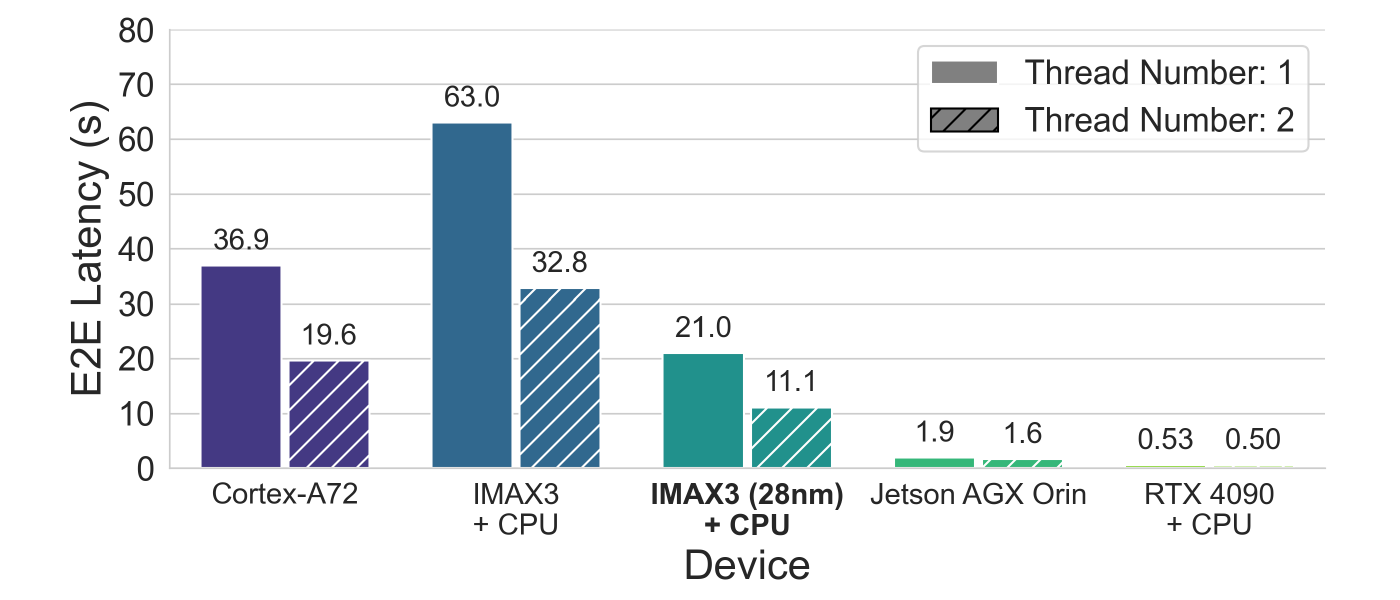}
      \subcaption{\small Whisper-tiny.en Q8\_0 model inference}
      \label{fig:e2e_Q80}
    \end{subfigure}
    
    \caption{\small E2E latency comparison by device. The IMAX (\SI{28}{\nm}) demonstrates a speedup over the CPU, while GPU remains the fastest.}
    \label{fig:e2e_latency}
  \end{figure}
\begin{figure}[t]
  \centering
  
  \begin{subfigure}{1.0\columnwidth}
    \centering
    \includegraphics[width=\textwidth]{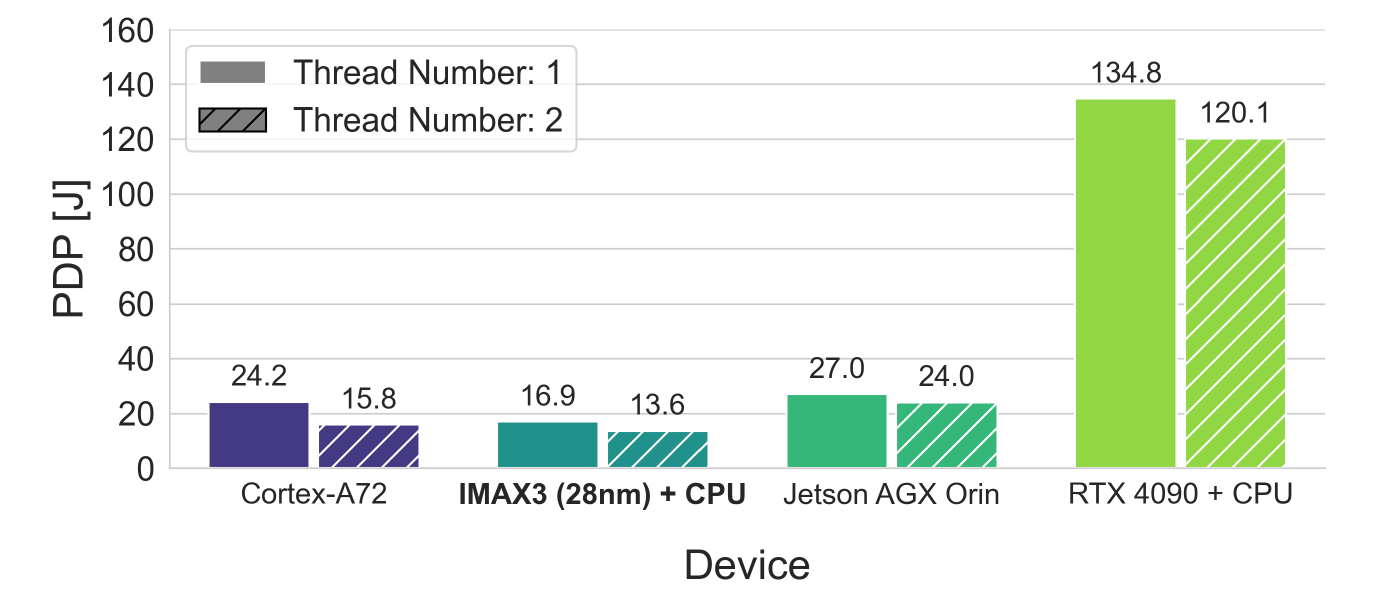}
    \subcaption{\small Whisper-tiny.en FP16 model inference}
    \label{fig:pdp_FP16} 
  \end{subfigure}
  
  \vspace{6pt} 
  
  \begin{subfigure}{1.0\columnwidth}
    \centering
    \includegraphics[width=\textwidth]{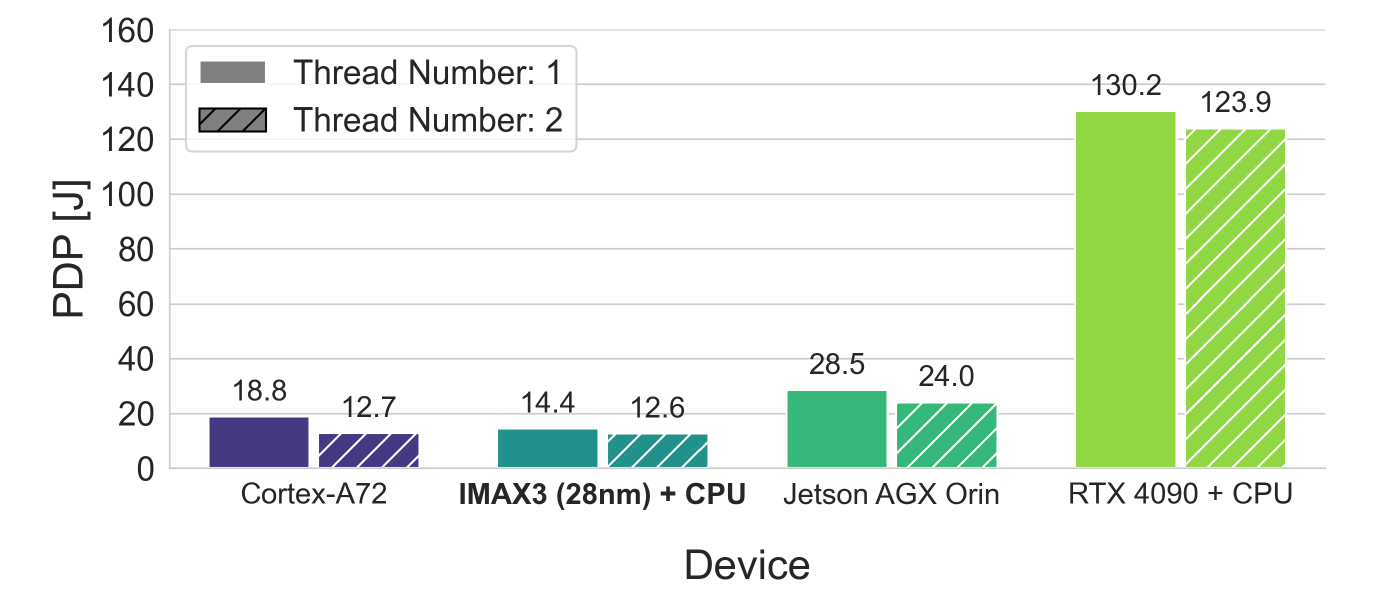}
    \subcaption{\small Whisper-tiny.en Q8\_0 model inference}
    \label{fig:pdp_Q80} 
  \end{subfigure}
  
  \caption{\small PDP performance comparison by device (lower is better). The IMAX (\SI{28}{\nm}) is more energy-efficient than other platforms.}
  \label{fig:pdp_comparison}
\end{figure}

\subsection{Evaluation of Processing Time and Power Consumption}
Fig.~\ref{fig:e2e_latency} shows the end-to-end latency for each platform. 
Fig.~\ref{fig:pdp_comparison} presents a comparison of the Power-Delay Product (PDP).
All evaluations were measured while varying the number of host CPU threads between one and two.

As shown in Fig.~\ref{fig:e2e_latency}, our IMAX with a 28\,nm ASIC projection recorded a latency of 13.5\,s for the FP16 model and 11.1\,s for the Q8\_0 model during two-thread execution. 
This performance does not match that of the high-end platforms. 
The Jetson AGX Orin edge GPU achieved 1.6\,s for both models, while the high-end RTX 4090 GPU reached 0.49\,s for FP16 and 0.50\,s for Q8\_0. 
This performance difference is attributable to the massive parallel computation resources of GPUs. 
However, in power-constrained edge applications, energy efficiency is often more important than absolute speed.
In this critical domain, the narrative shifts decisively in favor of IMAX.
The acceleration provided by IMAX is clear when compared to the host Cortex-A72 CPU alone, which required 24.4\,s for FP16 and 19.6\,s for Q8\_0.

Next, we evaluate the energy efficiency of each hardware platform. We adopt the PDP as our metric because it considers both execution time and power consumption. 
PDP is defined by Equation~\ref{eq:pdp}. 
\begin{equation}
  PDP = \text{Execution time} \times \text{Power consumption}
  \label{eq:pdp}
\end{equation}
A lower value signifies better energy efficiency, indicating the ability to complete a task faster with less total energy.
As Fig.~\ref{fig:pdp_comparison} illustrates, our IMAX ASIC projection demonstrates superiority in energy efficiency, a primary focus of this work. 
During two-thread execution, the PDP of IMAX for the FP16 model was 13.6\,J. 
This value is 1.76$\times$ better than the 24.0\,J of the Jetson AGX Orin and 8.83$\times$ better than the 120.1\,J of the RTX 4090. 
Furthermore, applying Q8\_0 quantization improved the IMAX PDP to 12.6\,J. 
This widened the efficiency gap, achieving 1.90$\times$ better efficiency than the Jetson AGX Orin and 9.83$\times$ better efficiency than the RTX 4090.

These results demonstrate that the IMAX architecture, optimized through the co-design detailed, has the potential to significantly outperform existing SOTA platforms in energy efficiency, 
even if it does not match the absolute processing speed of GPUs. 
This is a critically important characteristic for ASR applications on power-constrained edge devices.

\section{Discussion}
\label{discussion}
This section analyzes key aspects of our co-design to validate its effectiveness and provide architectural insights.
\begin{figure}[t]
    \centering

    \begin{subfigure}{1.0\columnwidth}
      \centering
      \includegraphics[width=\textwidth]{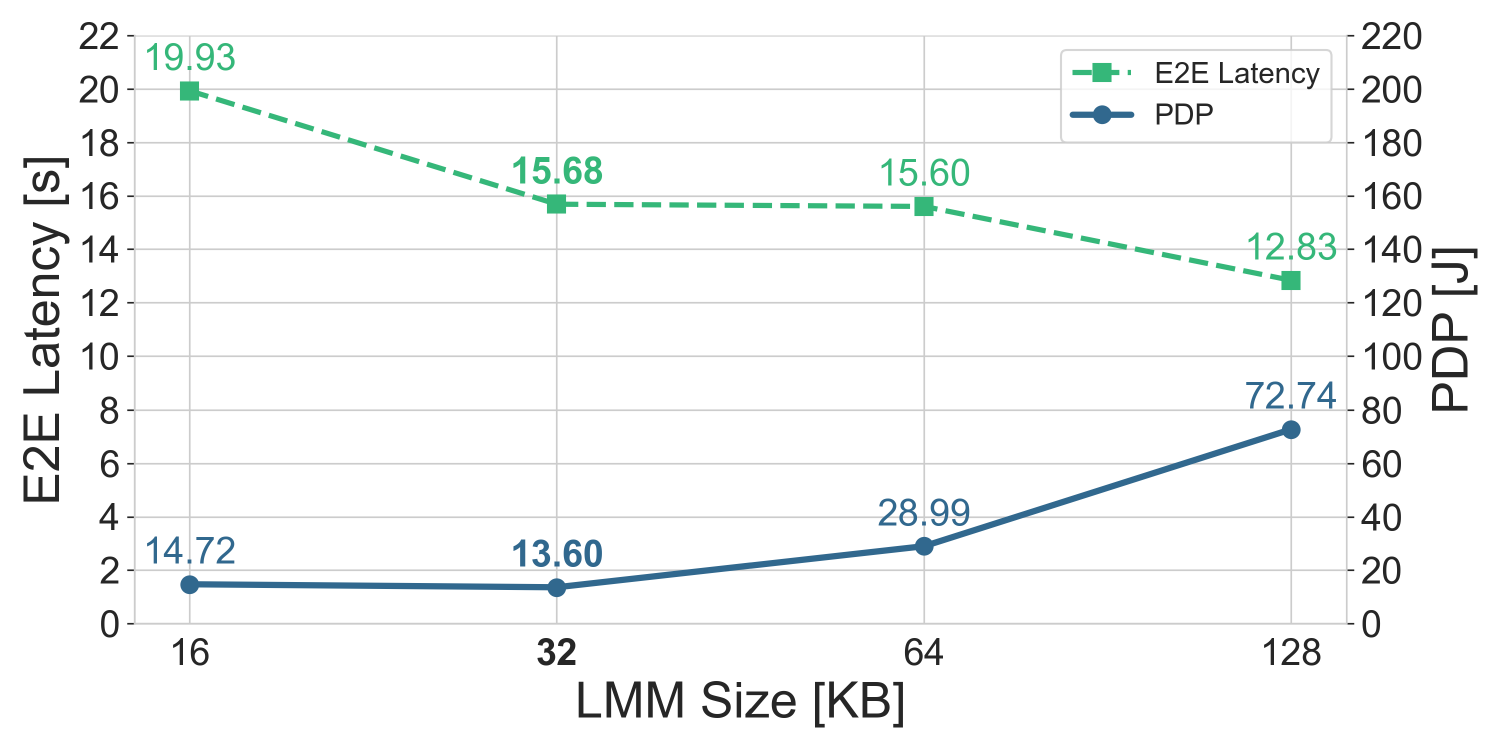}
      \subcaption{\small Whisper-tiny.en FP16 model inference.}
      \label{fig:e2e_latency_pdp_f16}
    \end{subfigure}

    \vspace{6pt} 

    \begin{subfigure}{1.0\columnwidth}
      \centering
      \includegraphics[width=\textwidth]{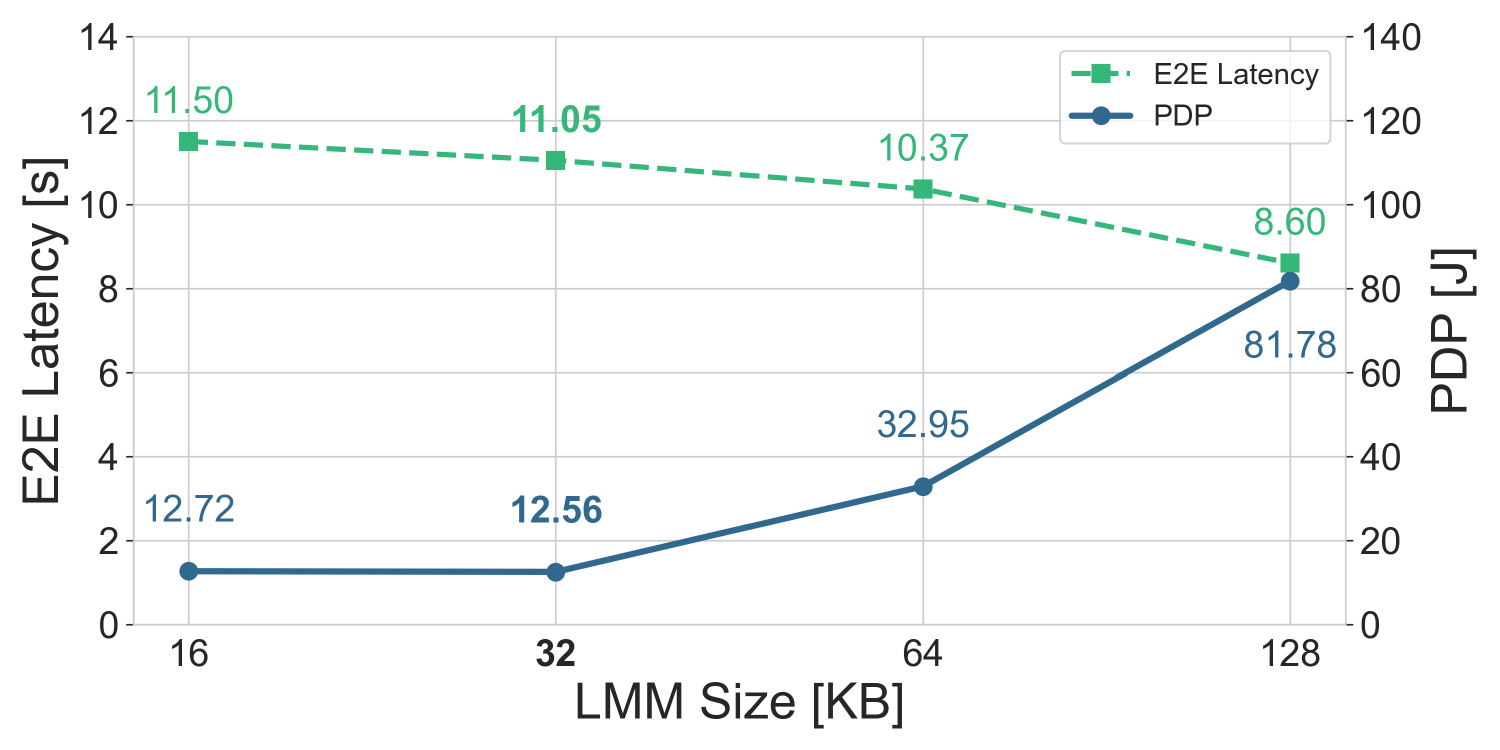}
      \subcaption{\small Whisper-tiny.en Q8\_0 model inference.}
      \label{fig:e2e_latency_pdp_Q80}
    \end{subfigure}

    \caption{\small E2E latency and PDP performance comparison by LMM size (lower is better). The minimum PDP point is 32\,KB for both FP16 and Q8\_0 models.}
    \label{fig:e2e_latency_pdp}
  \end{figure}

\subsection{Improving Energy Efficiency with LMM Size Optimization}
We validate our choice of a 32\,KB LMM. 
As shown in Fig.~\ref{fig:e2e_latency_pdp}, varying the LMM size reveals that the PDP reaches its minimum at 32\,KB for both FP16 and Q8\_0 models.
With a 16\,KB LMM, both latency and PDP degrade because the CPU must handle kernels that do not fit into memory. 
Conversely, increasing the LMM size to 64\,KB and 128\,KB provides only a marginal improvement in latency, which is outweighed by the increase in static power consumption, causing the PDP to steadily worsen.
This outcome demonstrates that simply increasing on-chip memory is not always better. Instead, it highlights the critical importance of finding an optimal point that considers both the application's memory footprint and the power characteristics. 
This analysis validates that our default 32\,KB LMM was the optimal choice for maximizing energy efficiency for the edge ASR task.

\subsection{ Verification of High Computational Efficiency}
\begin{figure}[t]
    \centering
    \includegraphics[width=1.0\columnwidth]{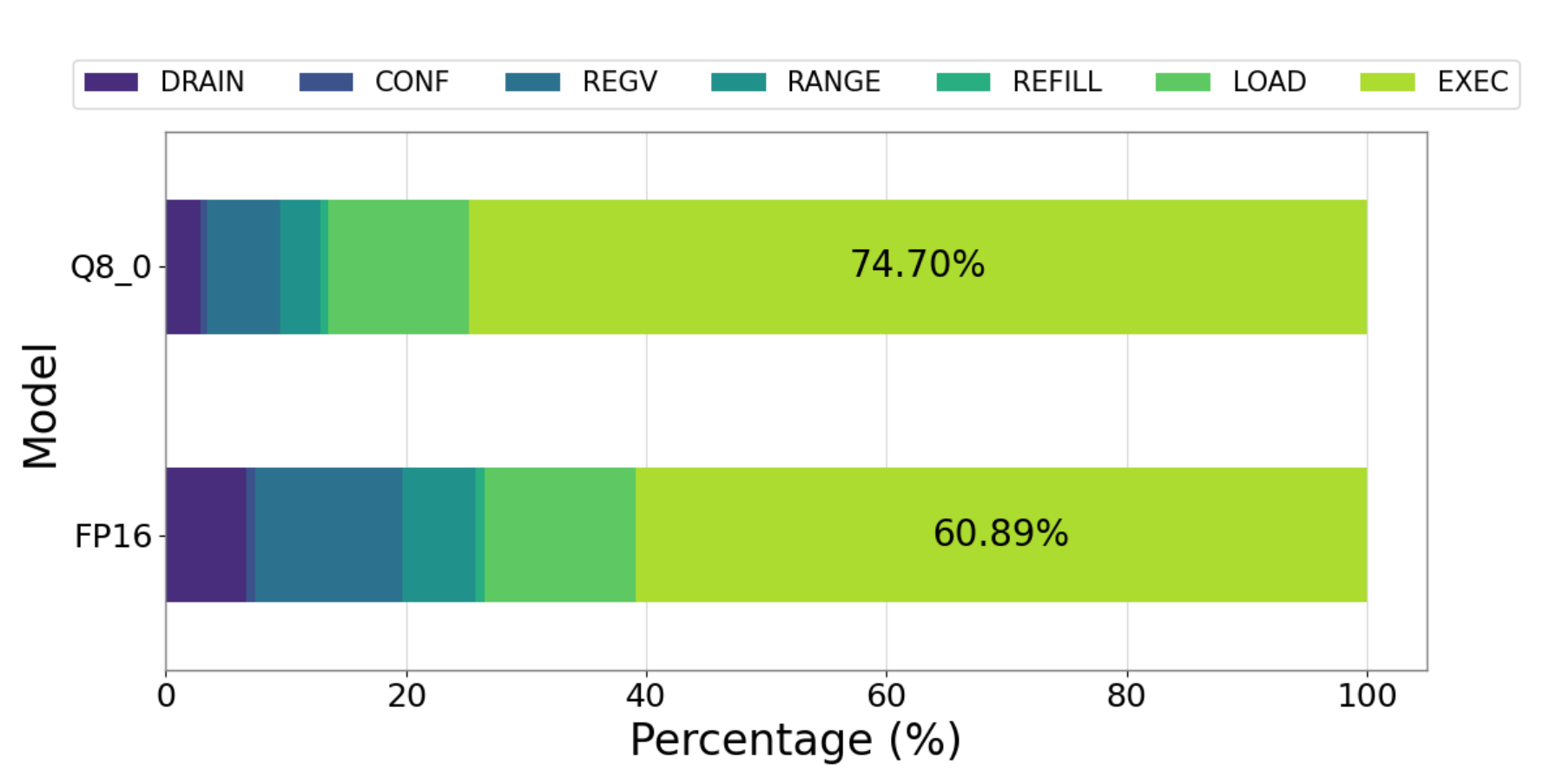}
    \caption{\small Execution time breakdown for IMAX kernels. The high ratio of EXEC indicates efficient, compute-bound operation.}
    \label{fig:comparison_stacked_bar_chart}
  \end{figure}

Next, we evaluate how effectively our co-design utilizes the computational resources of IMAX. 
Fig.~\ref{fig:comparison_stacked_bar_chart} presents a percentage breakdown of the kernel execution time components on IMAX under the optimized configuration. 
These components consist of EXEC (pure computation time on the PEs), LOAD/DRAIN (data transfer between DRAM and LMM), and CONF/REGV/RANGE/REFILL (IMAX configuration).
For the FP16 model, \SI{60.89}{\percent} of the total execution time is spent in the EXEC component, while for the Q8\_0 model, this figure rises to \SI{74.70}{\percent}. 
This high ratio of computation time is a noteworthy achievement. While many data-intensive AI workloads are memory-bound, causing processing units to stall, our results indicate that IMAX operates in a compute-bound state. 
Dedicating the majority of execution time to valuable computation suggests our co-design methodology has successfully mitigated data movement overheads.
This integration of kernel optimizations, efficient data handling, and LMM tuning effectively unlocks the high throughput potential of the IMAX architecture.


  \begin{table}[t]
    \centering
    \caption{Cumulative percentage of 3 edge ASR models up to the specified LMM size limit}
    \label{tab:lmm_size_base_and_small}
    \setlength{\tabcolsep}{4.8pt}
    \begin{tabular}{lcccccc}
      \toprule
      \textbf{Model} & \textbf{16\,KB} & \textbf{32\,KB} & \textbf{64\,KB} & \textbf{128\,KB} & \textbf{256\,KB} \\
      \midrule
      tiny\hspace{5pt}(\,\,\,78\,MB) & 66.35\,\% & \textbf{93.80\,\%} & 93.80\,\% &100.00\,\% & 100.00\,\% \\
      base\hspace{3pt}(148\,MB) & 66.55\,\% & 66.54\,\% &\textbf{94.17\,\%} & 97.08\,\% & 99.89\,\% \\
      small\hspace{0.2pt}(488\,MB) & 66.53\,\% & 66.52\,\% &\textbf{94.36\,\%} & 96.89\,\% & 99.89\,\% \\
      \bottomrule
    \end{tabular}
  \end{table}
\subsection{Scalability to Larger Models}
While this work focused on the Whisper-tiny.en model for edge devices, scaling our approach to larger models like base or small introduces new considerations. 
Larger models significantly increase the computational workload. 
For instance, the number of dot-product operations grows from 477,153 for the tiny model to 644,690 for base and 1,920,955 for small.
However, our analysis reveals that the memory footprint required for a single operation does not grow proportionally. 
This is a key advantage of our approach. 
Consequently, while the 32 KB LMM is insufficient for these larger models, forcing frequent and costly fallbacks to the CPU, a modest expansion of the LMM is highly effective. 
As shown in Table~\ref{tab:lmm_size_base_and_small}, increasing the LMM size to 64 KB would accommodate over \SI{94}{\percent} of the computational kernels for both the base and small models. 
This demonstrates that our architecture has high applicability to larger models without requiring excessive on-chip memory like 256 KB or 512 KB.

However, increasing the LMM to 64 KB raises static power consumption, as shown in Table~\ref{tab:lmm_size_power_simple}. 
Nevertheless, this power increase is likely an acceptable trade-off to avoid the severe performance degradation from CPU fallbacks. 
This result underscores the potential of our architecture and highlights that optimizing this power-performance balance through more sophisticated co-design will be a critical future task for efficiently handling even larger models.
\section{Conclusion}
\label{conclusions}
In this paper, we implemented and evaluated the Whisper ASR kernel on the IMAX CGLA architecture. 
Our results highlight a crucial design trade-off in modern AI accelerators. 
While our optimized IMAX configuration cannot match the low latency of high-end GPUs, it delivers vastly superior energy efficiency. 
Specifically, our \SI{28}{\nm} ASIC projection for the Q8\_0 model achieved a PDP of \SI{12.6}{\joule}, surpassing the NVIDIA Jetson AGX Orin and RTX 4090 by \num{1.90}$\times$ and \num{9.83}$\times$, respectively. 
This distinction is critical for our target domain of power-constrained edge applications, where battery life and thermal management are often more important design constraints than achieving the absolute lowest latency.

Future work will extend our approach to larger Whisper models, such as base and small, while also exploring further kernel optimizations like tuning the number of computational units. 
Through these efforts, we aim to advance the development of next-generation, energy-efficient AI accelerators.

\section*{Acknowledgment}

This work was supported by the JST-ALCA-Next Program (Grant Number JPMJAN23F4) and JSPS KAKENHI (Grant No. 22H00515). We also acknowledge the activities of VDEC, The University of Tokyo, in collaboration with NIHON SYNOPSYS G.K.

\bibliographystyle{IEEEtran}
\bibliography{bibliography}

\end{document}